\documentclass[a4paper,11pt]{article}
\pdfoutput=1 % if your are submitting a pdflatex (i.e. if you have
\usepackage{jheppub} % for details on the use of the package, please
\usepackage{lineno}

\usepackage[T1]{fontenc}

\usepackage{amsfonts}
\usepackage{placeins}
\usepackage{graphicx}

\usepackage{color,amsmath,amssymb,bm}
\usepackage{tensor}
\usepackage[normalem]{ulem}

\def\blfootnote{\xdef\@thefnmark{}\@footnotetext}
\renewcommand{\sout}{\bgroup \color{red} \ULdepth=-.5ex \ULset}

\usepackage{amsmath}    % need for subequations
\usepackage{graphicx}   % need for figures
\usepackage{color}
\usepackage{bm} % for bold caligraphic font
\usepackage{soul}
\usepackage{graphicx}
\usepackage{soul}
\usepackage{textcomp}
\usepackage{hyperref}
\usepackage[toc,page]{appendix}
\usepackage{ulem}
\usepackage{subcaption}
\usepackage{floatrow}

\def \be {\begin{equation} }
\def \ee {\end{equation}}
\def \bes {\begin{subequations} }
\def \ees {\end{subequations}}

\def \a {\alpha}
\def \b {\beta}

\def \d {\delta}
\def \e {\varepsilon}
\def \g {\gamma}
\def \k {\kappa}
\def \o {\omega}

\def \l {\lambda}

\def \s {\sigma}

\def \vp {\bm{p}}
\def \vq {\bm{q}}
\def \vx {\bm{x}}
\def \vy {\bm{y}}

\def \vo {\bm{\o}}

\def\vo{\bm{\o}}

\def \vv {\bm{v}}

\def \vgamma {\bm{\gamma}}
\def \hp {\hat{\vp}}

\def \le {\left}
\def \ri {\right}
\def \O {\Omega}

\def \<{\langle}
\def \>{\rangle}
\def \+{\dagger}

\def \[{\left[}
\def \]{\right]}

\def \pd {\partial}

\def \Tr {{\rm Tr}}
\def \no {\nonumber}
\def \CP {{\cal P}}

\def \CT {{\cal T}}

\def \sA {{\cal A}}

\newcommand{\beq}{\begin{equation}}
\newcommand{\eeq}{\end{equation}}
\newcommand{\bea}{\begin{eqnarray}}
\newcommand{\eea}{\end{eqnarray}}

\title{Spin polarization induced by the hydrodynamic gradients}
\author[a]{Shuai Y.\,F.~Liu}
\author[a,b]{and Yi Yin}

\affiliation[a]{Quark Matter Research Center, Institute of Modern Physics, Chinese Academy of Sciences, Lanzhou, Gansu, 730000, China}
\affiliation[b]{University of Chinese Academy of Sciences, Beijing, 100049, China}

\emailAdd{lshphy@gmail.com}
\emailAdd{yiyin@impcas.ac.cn}

\date{\today}
\abstract{
We systematically analyze the effects of the derivatives of the hydrodynamic fields on axial Wigner function that describes the spin polarization vector in phase space. We have included all possible first-order derivative contributions that are allowed by symmetry and compute the associated transport functions at one-loop using the linear response theory. In addition to reproducing known effects due to the temperature gradient and vorticity, we have identified a number of potentially significant contributions that are overlooked previously. In particular, we find that the shear strength, the symmetric and traceless part of the flow gradient, will induce a quadrupole for spin polarization in the phase space. 
We refer to this novel effect as the shear-induced polarization (SIP).
Our results, together with hydrodynamic gradients obtained from hydrodynamic simulations, can be employed as a basis for the interpretation of the $\Lambda$ (anti-$\Lambda$) spin polarization measurement in heavy-ion collisions.
}

\begin{document}
%\pacs{25.75.Dw, 12.38.Mh, 25.75.Nq}
%\keywords{Heavy-Flavor Transport, Quark-Gluon Plasma, Ultrarelativistic Heavy-Ion Collisions}
\maketitle
\flushbottom
%%%%%%%%%%%%%%%%%%%%%%%%%%%%%%%%%%%%%%%%%%%%%%%%%%%%%%%%
\section{Introduction}

The study of spin polarization of $\Lambda$ and $\bar{\Lambda}$ hyperons produced in heavy-ion collisions opens a new avenue to explore the properties of hot and dense QCD matter from the perspective of the spin degree of freedom. 
Extensive theoretical and phenomenological investigations are devoted to the effects of vorticity on spin polarization~\cite{Liang:2004ph,Becattini:2013fla,Becattini:2020sww} as well as the related transport phenomenon involving spin~\cite{Florkowski:2017ruc,Hattori:2019lfp,Liu:2019krs,Fukushima:2020ucl,Shi:2020htn,Li:2020eon,Singh:2020rht}.
STAR collaboration's measurement of global (phase space averaged) $\Lambda$ polarization agrees with theoretical prediction~\cite{STAR:2017ckg,Adam:2018ivw}.
However, 
when it comes to the phase space distribution of $\Lambda$ spin polarization~\cite{Niida:2018hfw,Adam:2019srw},
the observed quadrupole pattern is firmly in tension with the theories assuming such quadrupole pattern is solely induced by thermal vorticity~\cite{Becattini:2017gcx,Xia:2018tes},
 a specific linear combination of vorticity and temperature gradients.

Nevertheless, vorticity and temperature gradient are not the only examples of gradient, i.e., the derivatives of hydrodynamics field (or derivatives for short). 
More generally, 
one may ask if other derivatives, such as shear strength (the traceless and symmetric part of the flow gradient, see below), could induce spin polarization in the phase space. 
Therefore, to establish the physical interpretation of experimental results, it is desirable to first consider all possible effects that may arise from derivatives. 
Such a systematic analysis should be possible when the system's typical gradient is small compared with the mean free path.
This systematic analysis is what we shall do in this paper.
We shall consider the axial Wigner function in phase space $\sA^\mu$ which is closely related to the phase space distribution of spin vector.
Then, we expand $\sA^\mu$ in terms of the first order derivatives together with Lorentz tensors formed by single-particle momentum, and include all possible terms allowed by symmetry.

To determine transport functions coming with such expansion, 
we compute relevant correlation functions that contain the information about the response of spin polarization distribution to the derivatives. 
Alternatively, one can extract those coefficients by matching to the analysis using quantum kinetic theory with the spin degree of freedom~(see for example Refs.~\cite{Chen:2014cla,Chen:2015gta,Hattori:2019ahi}).
We shall see the agreement between both approaches for chiral fermions.

One remarkable effect uncovered from the present analysis is that the shear strength will induce quadrupole in the spin polarization distribution.
We shall refer such an effect as ``shear-induced polarization'' (SIP). 
\footnote{
In some context, the shear strength is refered as ``strain rate'', see Ref.~\cite{Crooker_2005,2005PhRvL..95j7203M} for a discussion of strain-induced spin current in the condensed matter system.
}
As we shall explain in detail in Sec.~\ref{sec:mag},
physics origin of SIP is closely related to the single-particle magnetization current term, the importance of which is well-known in chiral kinetic theory~\cite{Chen:2014cla,Kharzeev:2016sut}.
In many previous studies of spin polarization, the flow configuration with a vanishing shear strength is assumed. 
Nevertheless, 
sizable shear strength is undoubtedly present in heavy-ion collisions and may potentially lead to observable effects.

This paper is organized as follows. 
We present the derivative expansion for axial Wigner function in phase space in Sec.~\ref{sec:expansion}. In Sec.~\ref{sec:kubo}, we derive Kubo relation which relates transport functions coming with the expansion with relevant correlation functions, which we evaluate at one-loop in Sec.~\ref{sec:cal}.
We collect and interpret our results in Sec.~\ref{sec:dis}. 
Sec.~\ref{sec:outlook} is devoted to conclusion and outlook. 
We use the most minus signature $(+,-,-,-)$ and adopt the notation $ \epsilon^{0123}=-\epsilon_{0123}=1$.

\section{Derivative expansion} 
\label{sec:expansion}

We consider many-body systems whose degrees of freedom include fermion particles/anti-particles. 
We shall study the phase space density of axial current that are described by the following Wigner function:
\begin{align}
\label{sP-def}
\sA^{\mu}(t,\vx;\vp)&=\e_{\vp}\int d^{3}\vy\, e^{-i \vp\cdot \vy}\,
\langle
\hat{{\cal J}}^{\mu}_{5}(t,\vx;\vy)
\rangle\, , 
\end{align}
where we have included the single-particle energy $\e_{\vp}=\sqrt{\vp^{2}+m^{2}}$ in the definition of $\sA^{\mu}$ in Eq.~\eqref{sP-def} so that $\sA^{\mu}$ is an axial Lorentz four vector (see also Eq.~\eqref{J5} below). 
In Eq.~\eqref{sP-def},
the thermal ensemble average $\langle\ldots\rangle$ is taken over the operator
\begin{align}
\hat{{\cal J}}^{\mu}_{5}(t,\vx;\vy)\equiv \bar{\psi}(t,\vx-\frac{\vy}{2})\, \g^{\mu}\gamma^{5}\, \psi(t,\vx+\frac{\vy}{2})\, . 
\end{align}
Here, $\g^{5},\g^{\mu}$ denote the standard gamma matrices and $\psi$ represents a generic Dirac field. 
For simplicity, we shall assume there is only one ``flavor'' of fermions with an arbitrary mass $m$. 
Note the integration of $\sA^{\mu}$ over the Lorentz invariant phase space volume gives axial current:
\begin{align}
\label{J5}
J^{\mu}_{5}(t,\vx)=\langle \hat{{\cal J}}^{\mu}_{5}(t,\vx;\vy=0)\rangle=\frac{1}{(2\pi)^{3}\,}\int \frac{d^{3}\vp}{\e_{
\vp}}\, \sA^{\mu}(t,\vx,\vp)\, . 
\end{align}

Throughout this paper, we are working in the regime where the typical frequency and momentum, $q_{0},\vq$ associated with the system's inhomogeneity is much smaller than the characteristic energy of fermions/anti-fermions.
In this regime,
the transition between fermion states and anti-fermions states are suppressed, and we can have the division $\sA^{\mu}=\sA^{\mu}_{+}+\sA^{\mu}_{-}$ with $\sA^{\mu}_{+}$ and $\sA^{\mu}_{-}$ describing fermion and anti-fermion distribution respectively.
When the system is in thermal equilibrium and homogeneous, 
the $\sA^{\mu}_{\pm}$ should equal to zero.\footnote{
We have assumed that the equilibrium state under study is parity-invariant.
For a system with a non-zero axial density $n_{A}$, 
$\sA^{\mu}\propto n_{A} u^{\mu}$ is allowed even at zeroth order in derivative expansion.
}
Now we consider the situation in which slow varying hydrodynamic fields are present.
We shall assume that the system has a vector $U(1)$ symmetry and denote associated chemical potential by $\mu$. 
Consequently, 
the hydrodynamic fields are $u^{\mu}$, $T$ (or its inverse $\b$) and $\mu$. 
The presence of derivatives of hydrodynamic fields, or derivatives, in turn, should induce $\sA^{\mu}_{\pm}$. 
Our goal is to look for an expansion of $\sA^{\mu}$ up to to the first order in the derivatives.
The corrections to the resulting expression should be suppressed when $q_{0},q=|\vq|\ll \tau^{-1}_{R}$ where $\tau_{R}$ is the typical relaxation time.

Let us now collect the first order derivatives:
\bes
\label{gradient}
\begin{align}
&\,\theta= \pd_{\perp} \cdot u\, , 
\nonumber\\
&\,\o^{\mu}=\frac{1}{2}\epsilon^{\mu\nu\a\lambda}u_{\nu}\pd^{\perp}_{\a}u_{\lambda}\, , 
\qquad
\b^{-1}\pd^{\mu}_{\perp}\b\, , \qquad
\pd^{\mu}_{\perp}(\b \mu)\, , 
\nonumber\\
\label{shear}
&\,\s^{\mu\nu}=\frac{1}{2}(\pd^{\mu}_{\perp}u^{\nu}+\pd^{\nu}_{\perp}u^{\mu})-\frac{1}{3}\Delta^{\mu\nu}\theta \, .	
\end{align}
\ees
where  $\omega^\mu$ and $ \sigma^{\mu\nu} $ are the fluid vorticity and  shear strength, respectively. 
Following a standard notation for hydrodynamics, we have defined the project
\begin{align}
&\,\Delta^{\mu\nu}\equiv \eta^{\mu\nu}-u^{\mu}u^{\nu}\, ,
\end{align}
and denote the transverse part of a generic vector $V^{\mu}$ by $V^{\mu}_{\perp}=\Delta^{\mu\nu}V_{\nu}$. 
Note the terms involving time derivatives in the fluid rest frame, i.e $D=u\cdot \pd$, are not included above since they can be replaced by a specific combination of spatial gradient terms in Eq.~\eqref{gradient} using hydrodynamic equation of motion.

We have considered the most general expression of $\sA^{\mu}_{\pm}$ that can be constructed from  $u^{\mu}, \e^{\mu\nu\a\b}, \eta^{\mu\nu}, p^{\mu}, p^{\mu}p^{\nu},\ldots $ and derivatives listed in Eq.~\eqref{gradient}, and require the resulting expression to be consistent with symmetries, including discrete ones. 
Decomposing $\sA^{\mu}_{\pm}$ into longitudinal and transverse parts: $\sA^{\mu}_{\pm}=f_{A,\pm}u^{\mu}+A^{\mu}_{\perp,\pm}$,
we find (see Appendix.~\ref{sec:gradient-more} for details):
\begin{align}
\label{f-exp}
f_{A,\pm}&=\tilde{c}_{\o,\pm}\, v\,\cdot \o\,
\\
\label{A-exp}
\sA^{\mu}_{\perp,\pm}&= 
c_{\o,\pm}\,\o^{\mu}+c_{T,\pm}\,\epsilon^{\mu\nu\a\l} u_{\nu}\, v_{\a}\,\b^{-1}(\pd^{\perp}_{\l} \beta)+c_{\mu,\pm}\,\,\epsilon^{\mu\nu\a\l} u_{\nu} v_{\a}\le(\pd^{\perp}_{\l}(\b\mu)\ri)
\nonumber\\
&\, +g_{\s,\pm}\,\epsilon^{\mu\nu\l\a}u_{\nu}Q\indices{_{\l}^{\rho}} \,\s_{ \rho\a}+g_{\o,\pm}\,Q^{\mu\nu}\o_{\nu}
\end{align}
where we have defined the single-particle velocity and energy
\begin{eqnarray}
\label{v-def}
v^{\a}=\frac{p^{\a}_{\perp}}{\e_{u}}\, ,
\qquad
\e_{u}=p\cdot u\, ,
\end{eqnarray}
and the generalized quadrupole moment is given by 
\begin{eqnarray}
Q^{\mu\nu}=-\frac{p^{\mu}_{\perp}p^{\mu}_{\perp}}{p^{2}_{\perp}}+\frac{1}{3}\Delta^{\mu\nu}\, .
\end{eqnarray}
In the fluid rest frame, $Q^{ij}=\hp^{i}\hp^{j}-\frac{1}{3}\d^{ij}$.
\footnote{
We use $\d_{ij}$ to denote the standard Kronecker delta function, i.e. $\delta_{xx}=\delta_{yy}=\delta_{zz}$. 
Because of the most minus metric that we use, $\delta_{ij}=-\delta_{i}^{j}$ and in the fluid rest frame $\Delta_{ij}=-\delta_{ij}$.
}
Note for chiral fermions, $f_{A,\pm}/\e_{\vp}$ coincides with the distribution function of axial charge (c.f.~\eqref{J5}).

In Eqs.~\eqref{f-exp}, \eqref{A-exp}, those scalar functions 
\begin{align}
\label{coeff}
\tilde{c}_{\o,\pm},c_{\o,\pm}, c_{T,\pm},g_{\s,\pm},g_{\o,\pm}, c_{\mu,\pm}
\end{align}
 may be viewed as generalized transport coefficients which determine the magnitude of the induced axial Wigner function.
They are function of $T,\mu$ as well as the Lorentz scalar $\e_{u}$,
but we will keep those dependence implicit. 
To this point, those coefficients are generic and should be determined from microscopic calculations. 
We shall compute them at one-loop order, 
see Eqs.~\eqref{c-results}, \eqref{g-results}, \eqref{c-T} or the collection of results.

Let us interpret each individual terms in Eqs.~\eqref{f-exp}, \eqref{A-exp}.  
As we shall elaborate in Sec.~\ref{sec:mag}, Eq.~\eqref{f-exp} results from the energy-shift due to spin-vorticity coupling in chiral limit.
There are five terms shown in Eq.~\eqref{A-exp}. 
The first one tells us nothing but the familiar phenomenon that vorticity will induce the spin polarization. 
The second and third terms of Eq.~\eqref{A-exp} describe the thermally-induced spin-Hall effect with temperature and chemical gradient (thermodynamic force) playing the role of the analog of electric field, 
see Ref.~\cite{Liu:2020dxg} for further discussion. 
While the first line of Eq.~\eqref{A-exp} describes known effects of vorticity and temperature/chemical potential gradient to the axial current distribution, 
physics of the second line has not been fully appreciated to date.
It tells us that both velocity stress tensor $\s_{\mu\nu}$ and vorticity $\o^{\mu}$ could induce a quadrupole in phase space. 
Such flow-gradient-induced phase space quadrupole is allowed by symmetry, 
and hence the associated transport functions $g_{\o}, g_{\s}$ should be non-zero in general. 
Both terms contain only one spatial gradient,
and their effects can potentially be as significant as other terms in Eq.~\eqref{A-exp}.

Summarizing this section, 
we obtain the derivative expansion for axial Wigner function in phase space
Eqs.~\eqref{f-exp}, \eqref{A-exp}. 
If the system is isotropic in the absence of the derivatives, only the first term in Eq.~\eqref{A-exp} will survive after the phase space integration in axial current $J^{\mu}_5(t,\vx)$~\eqref{J5}.
However, physics is richer in phase space. 
According to Eq.~\eqref{A-exp}, different derivatives terms will induce different multipole moments respectively.
Our next goal is to evaluate associated transport functions listed in Eq.~\eqref{coeff}.

\section{Kubo relation}
\label{sec:kubo}

In this section, we shall establish Kubo relations which relate transport functions in Eqs.~\eqref{f-exp}, \eqref{A-exp} to the behavior of relevant correlation function under appropriate limit(s).
The method we shall use is inspired by the classical work of Luttinger~\cite{Luttinger1964}, but with a number of non-trivial generalization (see below). 
In Ref.~\cite{Liu:2020dxg}, 
we have computed $c_{T,\pm}, c_{\mu, \pm}$ at one-loop order using this formalism. 
In what follows, 
we shall focus on the Kubo relations for $\tilde{c}_{\omega}, c_{\o}$ and $g_{\s}, g_{\o}$. 
For this reason, we shall not include effects induced by the derivatives of $T$ and $\mu$ in this section.

Let us consider a fluid which is at rest $u^{\mu}=(1,{\bm 0})$ initially. 
Then, we turn on a metric perturbation of the form $g_{\mu\nu}=\eta_{\mu\nu}+h_{\mu\nu}$ with only $h_{0i}(t,\vx)$ different from zero.
The presence of $h_{0i}(t,\vx)$ will in turn induce flow $u^{\mu}=(1,{\bm u}(t,\vx))$.
To perform the linear response analysis, we count $h_{0i}$ and ${\bm u}$ to be ${\cal O}(\d)$ where $\d$ is a small parameter. 
In hydrodynamic regime when typical frequency $q_{0}$ and gradient $\vq$ associated with $h_{0i}(t,\vx)$ is much smaller than the relaxation rate $1/\tau_{R}$, 
we may expand $\sA^{i}_{\perp},f_{A}$ to the first order in both gradient and $\d $: 
\bes
\label{A-exp-3d}
\begin{align}
f_{A,\pm}&=\tilde{c}_{\o,\pm}\, v^{j}\o_j\,+\tilde{c}^h_{\o,\pm}\, v^{j}B^h_j\nonumber\\
\sA^{i}_{\pm}&=\le( c_{\o,\pm}\d^{ij}+g_{\o,\pm}\, Q^{ij}\ri) \o_{j}-g_{\s,\pm}\,\epsilon^{ilm}Q_{lj}\s^{j}_{\,\,m}
 \\
&\,+ \le( c^{h}_{\o,\pm}\d^{ij}+g^{h}_{\o,\pm}\, Q^{ij}\ri) B^{h}_{j}-g^{h}_{\s,\pm}\,\epsilon^{ilm}Q_{lj}(\s^{h})^{j}_{\,\,m}\, ,
\end{align}
\ees
where $\vv=\vp/\e_{\vp}$ and the summation over the dummy index is understood. %red{?where is the label for $f_A$}
In this section, we use $\s^{ij}$ (and similar for $\o^{\mu},Q^{\mu\nu}$) to denote $i,j$ components of $\s^{\mu\nu}$ defined in Eq.~\eqref{shear} up to first order in $\d$\footnote{We use
$\epsilon^{ijk}=\epsilon^{0ijk}=-\epsilon_{ijk}$ to denote Levi-Civita symbol in 3d. 
},
i.e.,
\begin{align}
\label{du-def}
&\, \o^{i}=-\frac{1}{2}\epsilon^{ijl}\pd_{j}u_{l}\, , 
\qquad
\s^{i}_{\,\, j}=\frac{1}{2}\,\le(\pd^{i}u_{j}+\pd^{j}u_{i}+\frac{2}{3}\d^{i}_{\,\, j}(\pd^{l}u_{l})\,\ri)\, . 
\\
\label{vQ3d-def}
&\, Q_{ij}=\frac{p_ip_j}{p^2}-\frac{1}{3}\delta_{ij}\, .
\end{align}
We further decompose the gradient of $h_{0i}$ into symmetric and anti-symmetric part and define in analogous to Eq.~\eqref{du-def} that
\begin{align}
\label{s-h-def}
B^{i,h}=-\frac{1}{2}\epsilon^{ijl}\pd_{j}h_{0l}\, , 
\qquad
(\s^{h})^{i}_{\,\, j}=\le(\frac{1}{2}\,\le(\pd^{i}h_{0j}+\pd^{j}h_{0i}\ri)+\frac{1}{3} \d^{i}_{\,\, j}\,(\pd^{l}h_{0l})\ri)\, 
\end{align}
Note in the absence of the metric perturbation, 
Eq.~\eqref{A-exp-3d} matches with Eq.~\eqref{A-exp} as it should be.

If $\sA^{\mu}$ were covariant under local Lorentz transformation, 
its dependence on the metric is simply given by taking all derivatives in Eq.~\eqref{A-exp} as covariant ones.
However, 
$\sA^{\mu}$ is only covariant under global Lorentz transformation but not covariant \textit{local} Lorentz transformation when a metric perturbation is present.
That is why $c^{h}_{\o,\pm}, g^{h}_{\o,\pm},g^{h}_{\s,\pm}$ in Eq.~\eqref{A-exp-3d} are independent and can not be fixed by imposing local Lorentz covariance. 
In principle
one can embed the appropriate Gauge link to obtain the covariant Wigner function $\sA'^{\mu}$, 
see Refs.~\cite{Liu:2020flb,Hayata:2020sqz} for further discussion. 
However, 
for the purpose of determining $c_{\o},g_{\o}$ and $g_{\s}$, considering $\sA$ is sufficient since the difference between $\sA$ and the covariant one $\sA'$ should only appear at higher order in derivatives.

Next, 
we consider the linear response theory which tells us that the induced axial current distribution up to the first order in metric perturbation (in Fourier space) is given by (c.f~Eq.~\eqref{sP-def}):
\begin{align}
\label{A-response}
\sA^{\mu}_{\pm}(q_{0},\vq; \vp) = \e_{\vp}\,G^{\mu,0j}_{\pm}(q_{0},\vq; \vp)\, h_{0j}(q_{0},\vq; \vp)\, . 
\end{align}
where the retarded correlation function of $\hat{\sA}^{\mu}$ and stress-energy tensor operator $\hat{T}^{\mu\nu}$ reads:
\begin{eqnarray}
\label{G-def}
 &\,&  G^{\mu,0\nu}(t,\vx;\vy)=-i\, \langle[ \hat{{\cal J}}^{\mu}_{5}(t,\vx;\vy), \hat{T}^{0\nu}(0,0;0) ]\rangle \theta(t)\, ,
\end{eqnarray}
In Eq.~\eqref{A-response}, 
we have assumed that the correlation function can be divided into two parts,
\begin{align}
\label{G-decomp}
&G^{\mu,0\nu}(q_{0},\vq; \vp)=G^{\mu,0\nu}_{+}(q_{0},\vq; \vp)+G^{\mu,0\nu}_{-}(q_{0},\vq; \vp)\, ,
\end{align}
which describes the induced axial Wigner function of fermion and anti-fermion respectively. 
We shall show how to make such division explicitly in Sec.~\ref{sec:cal}.

To extract the desired transport functions from Eq.~\eqref{A-response}, in what follows, we shall consider the behavior of $\sA^{\mu}_{\pm}$ and the retarded correlation function under two different limits.
We shall detail the analysis of $\sA^{i}$. That of $f_{A}$ follows similar steps.

We first take the "slow limit" by which we mean:%$q_{0}\ll q$ while $q_{0},q\ll \tau^{-1}_{R}$ and denote this limit by $\lim\limits_{\textbf{s}}$:
\begin{align}
\label{lim-s}
\lim\limits_{\textbf{s}}: q_{0}\ll q\, , \textrm{and}\, q_{0},q\ll \tau^{-1}_{R}\, . 
\end{align}
In this limit, the flow induced by the metric perturbation is fully developed, 
and we would expect that the frame where the local fluid is at rest coincides with that of metric perturbation, i.e.,
\begin{align}
\lim_{s}\, \le(u_{j}-h_{0j}\ri)=0\, .
\end{align}
Taking the limit, Eq.~\eqref{A-exp-3d} becomes 
\begin{align}
\label{A-slow}
\lim\limits_{\textbf{s}}\sA^{i}_{\pm}&=\,[ (c_{\o,\pm}+c^{h}_{\o,\pm})\d^{il}+(g_{\o,\pm}+g^{h}_{\o,\pm})\, Q^{il}] B^{h}_{l}-(g_{\s,\pm}+g^{h}_{\s,\pm})\,\epsilon^{ilm}Q_{lj}(\s^{h})^{j}_{\,\, m}\, .
\end{align}
When taking the slow limit, we require $q_{0}/q \ll 1$ but $q_{0}$ is not necessarily zero. 
That is why  in Eq.~\eqref{A-slow} we can keep a non-zero shear strength $\s_{ij}$  which would excite only shear modes with a typical frequency of the order $q_{0}\sim\nu q^{2}\ll q$ where $\nu$ denotes the specific viscosity. 
Note in hydrodynamic regime $\nu q\ll 1$.

We now turn to the second limit, the fast limit, that : 
\begin{align}
&\,\lim\limits_{\textbf{f}}:q_{0}\gg q\, \,\textrm{and}\, q,q_{0}\, \,\gg \tau^{-1}_{R} \, . 
\end{align}
Note when taking the both limit, we still require that $\e_{0}\gg q_{0},q$ as we mentioned at the beginning of this paper. 
We shall further assume $\e_{0}\gg \tau^{-1}_{R}$ throughout, the assumption of which is justified when the interaction is weak.
In this limit, there is not enough time for the system to develop the flow velocity and hence flow velocity simply vanishes. 
However, we might still express $\sA$ in terms of gradient of metric:
\begin{align}
\label{A-map-f}
&\,\lim\limits_{\textbf{f}}\sA^{i}_{\pm}=\, \le( c^{h}_{\o,\pm}\d^{ij}+g^{h}_{\o,\pm}\, Q^{ij}\ri) B^{h}_{j}-g^{h}_{\s,\pm}\,\epsilon^{ilm}Q_{lj}(\s^{h})^{j}_{\,\, m}\,
\end{align}
One important assumption we made here is that $c^{h}_{\o},g^{h}_{\s}, g^{h}_{\o}$ are the same as those in Eq.~\eqref{A-exp-3d}. 
We can argue for this assumption as follows. 
The small parameter associated with the expansion in terms of the gradient of metric is $q_{0}/\e_{0}, q/\e_{0}$ where $\e_{0}$ denotes the typical single-particle energy.
Therefore the expansion coefficient should not be sensitive to the hierarchy between $q_{0},q$ and $\tau^{-1}_{R}$. 
Introducing the notation
\begin{align}
\lim_{s-f}(\ldots)\equiv 
\lim_{s}\,(\ldots) - \lim_{f}\,(\ldots)
\end{align}
where $\ldots$ represents an arbitrary function function of $q_{0},q$, 
we then have from Eqs.~\eqref{A-slow}, \eqref{A-map-f} that
\begin{align}
\label{A-c}
\lim_{s-f}\sA^{i}_{\pm}\,
=\le( c_{\o,\pm}\d^{ij}+g_{\o,\pm}\, Q^{ij}\ri) B^{h}_{j}-g_{\s,\pm}\,\epsilon^{ilm}Q_{lj}(\s^{h})^{j}_{\,\, m}\, . 
\end{align}
Comparing Eq.~\eqref{A-c} with Eq.~\eqref{A-response}, 
we may parametrize the behavior of $G$ as
\begin{align}
\label{G-lim}
\e_{\vp}\,\lim_{s-f} G^{i,0j}_{\pm}(q_{0},\vq; \vp) 
=\kappa^{ijm}_{\pm}(\vp) q_{m}\, , 
\end{align}
where the function $\kappa$ is given by:
\begin{align}
\label{kappa}
\kappa^{ijm}_{\pm}(\vp)=-\kappa_{ijm,\pm}(\vp)=\varepsilon_{\vp}\frac{\partial}{\partial q_m}\lim\limits_{\textbf{s}-\textbf{f}}G^{i,0j}_{\pm}(q_{0},\vq; \vp)\, . 
\end{align}
Comparing Eq.~\eqref{A-response} with Eq.~\eqref{A-c},
we further obtain the Kubo relation
\bes
\label{Kubo-c}
\begin{align}
\label{c-o}
&c_{\o\pm}=\frac{1}{3}i\,\varepsilon_{\vp}\epsilon^{ijm}
\,\kappa_{ijm}^{\pm}
\\
\label{g-o}
&g_{\o\pm}=\frac{3}{2}i\,\varepsilon_{\vp}\epsilon^{jlm}Q_{l}^{i}\,
 \kappa_{ijm}^{\pm}
\\
\label{g-s}
&g_{\s\pm}=i\,\varepsilon_{\vp}\,\epsilon^{il(j}Q^{m)}_l\, \kappa_{ijm}^{\pm}
\end{align}
where we have used the identity $\delta_{ii}=3, Q_{ij}Q^{ij}=2/3, \epsilon^{iml}Q_{l}^{\,\, j}\epsilon_{ki(m}Q_{j)}^{\,\,\,\, k}=1$,
and the brackets around a pair of indices, $(ij)$ mean that indices to be symmetrized.
By analyzing $f_{A}$, we obtain the following expression:
\begin{align}
\label{ct-o}
&\tilde{c}_{\o\pm}=-\frac{1}{\vv^2}\varepsilon_{\vp}\lim\limits_{\textbf{s}-\textbf{f}}\epsilon_{lmj}v^{l}\frac{i\partial}{\partial q_m}G^{0,0j}_{\pm}(q_{0},\vq; \vp)\, .
\end{align}
\ees
In the fast limit when the flow velocity is yet to develop,
$\sA^{i}_{\pm}$ solely comes from metric perturbation.
In the slow limit, $\sA^{i}_{\pm}$ contains the response from both the metric perturbation and flow gradient developed subsequently. 
Therefore one should subtract the $\lim\limits_{\textbf{f}}\sA^{i}_{\pm}$ from $\lim\limits_{\textbf{s}}\sA^{i}_{\pm}$ in order to determine the response of $\sA^{i}$ to flow gradient. 
For this reason, 
the transport functions, which describe the response to flow gradient, is determined by the difference between the behavior of correlation functions in slow and fast limit in the present Kubo relation .

We shall use Eq.~\eqref{Kubo-c} and Eq.~\eqref{ct-o} to compute $c_{\o},\,g_{\o},\,g_{\s}$ and $\tilde{c}_{\o}$ respectively in the subsequent section. 
We shall test our prescription by comparing the results at one-loop with those obtained from the quantum kinetic theory.

\section{One loop}
\label{sec:cal}

In this section, 
we shall explicitly evaluate the correlation functions defined in Eq.~\eqref{G-def} at one-loop.
We shall first present the calculation of $G^{i,0j}_{\pm}$ in details, 
and just state the final result for $G^{0,0i}_{\pm}$.
Note, since we are interested in the induced momentum space distribution, we shall only sum over loop frequency, but shall not integrate over the loop momentum $\vp$.
In Refs.~\cite{Landsteiner:2011iq,Lin:2018aon}, $\int d^3{\vp}/(2\pi)^3\, G^{i,0j}(q_0,\vq;\vp)$ is computed to study chiral vortical effect (CVE).

%%%%%%%%%%%%%%%%%%%%%%%%%%%%%%%%%%%%%%%%%%%%%%%%%%%%%%%%%%%%%%%%
%
%%%%%%%%%%%%%%%%%%%%%%%%%%%%%%%%%%%%%%%%%%%%%%%%%%%%%%%%%%%%%%%%
\begin{figure}[t]
	\centering
	\includegraphics[width=0.35\textwidth]{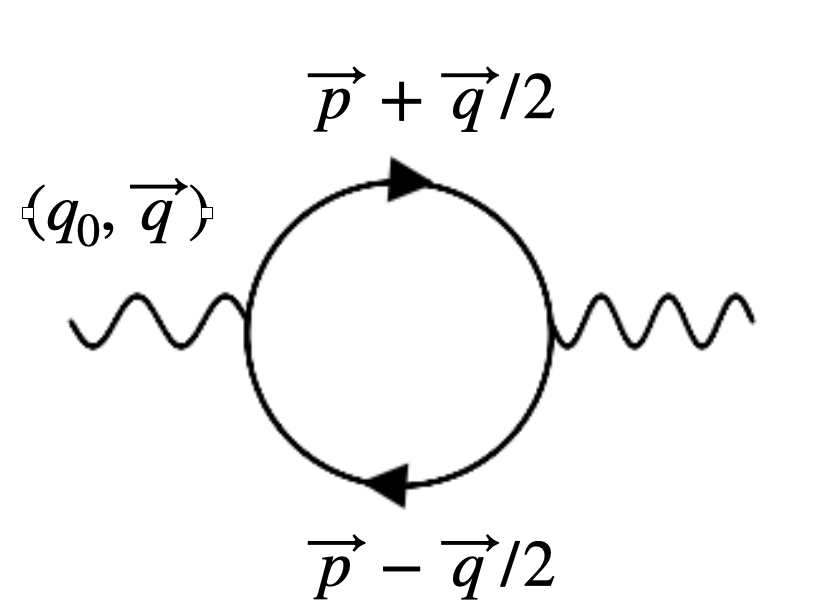}
	\caption{
		One loop diagram contributing to $G^{i,0\mu}$ defined in Eq.~\eqref{G-def}. Note, since we are interested in the induced momentum space distribution, we shall only sum over loop frequency (Matsubara frequency), but shall not integrate over the loop momentum $\vp$.
			}
	\label{fig:loop}
\end{figure}

We begin with the (symmetric) energy momentum tensor $\hat{T}^{\mu\nu}$ in terms of the Dirac field:
\begin{align}
\label{Tmunu}
\hat{T}^{\mu\nu}=\frac{i}{4}\,\bar{\psi} (\g^{\mu}\overleftrightarrow{\pd}^{\nu}+\g^{\nu}\overleftrightarrow{\pd}^{\mu})\psi \, .
\end{align}
where $\overleftrightarrow{\partial_\mu}$ is defined by $ A\overleftrightarrow{\partial_\mu} B=A\partial_\mu B-(\partial_\mu A)B $ for arbitrary function $ A $ and $ B $. 
Note by definition, $\hat{T}^{\mu\nu}$ is symmetric. 
Since $ \hat{T}^{\mu\nu}=\hat{T}^{\mu\nu}_{1}+\hat{T}^{\mu\nu}_{2}$ contains two terms where 
$\hat{T}^{\mu\nu}_{1}\equiv (i/4)\,\bar{\psi} \g^{j}\overleftrightarrow{\pd}^{0}\psi, \hat{T}^{\mu\nu}_{2}\equiv(i/4)\,\bar{\psi} \g^{0}\overleftrightarrow{\pd}^{j}\psi $, 
we divide correlation function into two parts accordingly:
\begin{align}
G^{\mu,0j} = G^{\mu,0j}_{1}+G^{\mu,0j}_{2}\, , 
\end{align}
where the $G^{\mu,0j}_{1,2}\sim\langle \le [ \hat{{\cal J}}^{\mu}_{5}, \hat{T}^{0j}_{1,2}\ri]\rangle$.
Introducing the Euclidean propagator as function of the Fermionic Matsubara frequency $\nu_{n}=\pi T(2n+1)+\mu$ and momentum $\vp$
\begin{eqnarray}
&\, S(i\nu_{n},\vp)
= \sum_{s=\pm}\,\Lambda_{s}(\vp)\,  \Delta_{s}(i\nu_{n},\vp)\, , 
\\
&\, \Lambda_{s}(\vp)= s\g^{0}\e_{\vp}-\vp\cdot \vgamma +m
\\
&\,\Delta_{s}(i\nu_{n},\vp)= (\frac{s}{2\e_{\vp}})\, \frac{1}{i \nu_{n}-s\e_{\vp}}\, ,
\end{eqnarray}
we can explicitly express correlation function in terms of (Matsubara) frequency and momentum space as
\begin{align}
\label{Gi0j-1}
\, G^{i,0j}_{1}(i\tilde{\o}_{n},\vq,\vp)
&=
\frac{1}{2\b}\, \sum_{\nu_{n}}\, \Tr
\left[
\g^{i}\g^{5}
S\left(i \nu_{n}+i\tilde{\o}_{n},\vp_{1}\right)\, \g^{j}\,(i \nu_{n}+\frac{i\tilde{\o}_{n}}{2})\,S(i\nu_{n},\vp_{2})
\right]
\\
\label{Gi0j-2}
\, G^{i,0j}_{2}(i\tilde{\o}_{n},\vq,\vp)
&=
\frac{1}{2\b}\, \sum_{\nu_{n}}\, \Tr
\left[
\g^{i}\g^{5}
S\left(i \nu_{n}+i\tilde{\o}_{n},\vp_{1}\right)\, \g^{0} \frac{p^{j}_{1}+p^{j}_{2}}{2}\,S(i\nu_{n},\vp_{2})
\right]\, 
\\
\label{G00j-1}
\, G^{0,0j}(i\tilde{\o}_{n},\vq,\vp)&=G^{0,0j}_{1}(i\tilde{\o}_{n},\vq,\vp)
\no\\
&=\frac{1}{2\b}\, \sum_{\nu_{n}}\, \Tr
\left[
\g^{0}\g^{5}
S\left(i \nu_{n}+i\tilde{\o}_{n},\vp_{1}\right)\, \g^{j}\,(i \nu_{n}+\frac{i\tilde{\o}_{n}}{2})\,S(i\nu_{n},\vp_{2})
\right]\,.
\end{align}
Note $\, G^{0,0j}_{2}=0$ due to the trace of the product of the associated gamma matrices vanishes.
Here, correlation function depends on Bosonic Matsubara frequency $\tilde{\o}_{n}=2n\pi T$ and we have defined 
$\vp_{1}= \vp+\frac{\vq}{2},\vp_{2}= \vp-\frac{\vq}{2}$.

To evaluate correlation functions, 
we need to take the trace
\begin{align}
\label{A-Trace}
&\, 
X^{ij}_{ss'}(\vp_{1},\vp_{2})\equiv
\Tr
\left[
\g^{i}\g^{5}
\Lambda_{s}(\vp_{1})\, \g^{j}\,\Lambda_{s'}(\vp_{2})
\right]
=-4i\epsilon^{ijm}\, \left(
s\e_{1}p_{2}-s'\e_{2}p_{1}\right)_{m}
\\
\label{B-Trace}
&\, Y^{i}(\vp_{1},\vp_{2})\equiv\Tr
\left[
\g^{i}\g^{5}
\Lambda_{s}(\vp_{1})\, \g^{0}\,\Lambda_{s'}(\vp_{2})
\right]
=4i\epsilon^{ijm}\,q_{j}p_{m}\, ,
\end{align}
and perform the summation over Matsubara frequency:
\bes
\label{M-sum}
\begin{align}
\label{M-sum-1}
I_{ss'}(q_0,\e_{1},\e_{2})&=T\sum_{\nu_{n}} \Delta_{s}(i\nu_{n}+i\tilde{\o}_{n},\vp_{1})\Delta_{s'}(i\nu_{n},\vp_{2}))|_{i\omega_n\rightarrow q_0+i0^{+}}
\nonumber\\
&=\sum_{ss'}\left(
\frac{-s s'}{4\e_{1}\e_{2}}
\right) \le(\frac{sn_{s}(\e_{1})-s'n_{s'}(\e_{2})}{q_0-s\e_{1}+s'\e_{2}+i0^{+}}\ri)
\\
\label{M-sum-2}
J_{ss'}(q_0,\e_{1},\e_{2})&=T\sum_{\nu_{n}} (i\nu_{n}+\frac{i\tilde{\o}_{n}}{2})\Delta_{s}(i\nu_{n}+i\tilde{\o}_{n},\vp_{1})\Delta_{s'}(i\nu_{n},\vp_{2})|_{i\omega_n\rightarrow q_0+i0^{+}}
\nonumber\\
&=(s \e_{1}- \frac{q_0}{2})I_{ss'}(q_0,\e_{1},\e_{2})+\frac{s}{4\e_{1}\e_{2}}\,n_{s'}(\e_{2}),
\end{align}
\ees
To obtain the real time correlation function, we analytically continue the Euclidean correlator by employing the replacement $i\omega_n\rightarrow q_0+i0^{+}$ in Eq.~\eqref{M-sum}.
Note, $0^{+}$ here may be understood as the inverse of relaxation time $\tau^{-1}_{R}$.
Here,  $\e_{1,2}=\sqrt{m^{2}+\vp^{2}_{1,2}}$ and $ n_{\pm}(\e)= 1/(e^{\beta (\e\mp\mu)}+1)$ is the Fermi Dirac distribution for particle ($ n_{+} $) and antiparticle ($ n_{-} $) with chemical potential $ \mu $. 
To obtain the above expression, we use $ n_{+}(s\varepsilon)\rightarrow s\,n_{s}(\varepsilon)$ and drop the vacuum contribution ``1" in the relation $ n_{+}(-\varepsilon)=1-\,n_{-}(\varepsilon) $.

We now express the retarded correlation functions  Eq.~\eqref{Gi0j-1}$\sim$\eqref{G00j-1} as
\begin{align}
\label{G1-in-J}
&\, G^{i,0j}_{1}(q_0,\vq,\vp)=\frac{1}{2}\sum_{ss'}\, X^{ij}_{ss'}(\vp_{1},\vp_{2})\, J_{ss'}(q_{0},\e_{1},\e_{2})\, , 
\\
\label{G2-in-J}
&\, G^{i,0j}_{2}(q_0,\vq,\vp)=\frac{1}{2}\,Y^{i}(\vp_{1},\vp_{2})\,p^{j}\,\sum_{ss'}\, \, I_{ss'}(q_{0},\e_{1},\e_{2})\, , 
\\
\label{G0-in-J}
&\, G^{0,0j}_{1}(q_{0},\vq,\vp)
=-\frac{1}{2}\,Y^{j}(\vp_{1},\vp_{2})\,\sum_{ss'} J_{ss'}(q_{0},\e_{1},\e_{2})
\end{align}
To proceed, we set up the power counting
\begin{align}
\frac{q_{0}}{\e_{0}}, \frac{q}{\e_{0}}={\cal O}(\d_{1})\,
\end{align}
where $\d_{1}$ is a small parameter. 
Summing up the contribution from $G_{1}$ and $G_{2}$ and expanding the result to the first order of $\d_{1}$ yield the expression
\begin{align}
\label{G-i0j-final}
G^{i,0j}_{\pm}(q_{0},\vq,\vp)&= -\frac{i}{2}\epsilon^{ijm}q_{m}\, \le[(\mp\Delta \e D_{\pm})-b_{\pm}\ri]
-\frac{i}{2}\epsilon^{ijl}v_{l}\Delta \e\, q_{0}D_{\pm}
\no\\
&-\frac{i}{2}\epsilon^{iml}q_{m}v_{l}v^{j}\le[(\mp\Delta \e D_{\pm})+b_{\pm}\ri]
\, \\
\label{G-00j-final}
G^{0,0j}_{\pm}(q_{0},\vq,\vp)&=
\pm\frac{i}{2}\epsilon^{jml}q_{m} v_{l}\, (\mp\Delta \e D_{\pm}).
\end{align}
The first line of Eq.~\eqref{G-i0j-final} is from $ G^{i,0j}_{1,\pm} $ and the second line of Eq.~\eqref{G-i0j-final} is from $ G^{i,0j}_{2,\pm}$.  
For notational brevity, we have defined
\begin{align}
\label{ab_defines}
&\,\vv = \frac{\vp}{\e_{\vp}}\,, 
\qquad
\Delta \e = \vv\cdot \vq=-v^m q_m\,,
\qquad
D_{\pm}\equiv \frac{a_{\pm}}{q_{0}\mp\Delta\e+i 0^{+}}
\\
&\,a_{\pm}=-\frac{\pd n_{\pm}(\e_{\vp})}{\pd \e_{\vp}}=\b\,n_{\pm}(\e_{\vp})\le(1-n_{\pm}(\e_{\vp})\ri)\,,
\qquad
b_{\pm}\equiv \frac{n_{\pm}(\e_{\vp})}{\e_{\vp}}
\,. 
\end{align}

As we discussed earlier, we need to decompose $G^{i,0j}$ (and similarly for other correlation functions) into two parts, $G^{i,0j}_{\pm}$ (see  Eq.~\eqref{G-decomp}), 
so that the induced axial current distribution for fermions and anti-fermions are given by Eq.~\eqref{A-response}.
Observing that $\sA_{+}$ ($\sA_{-}$) should only depend on $a_{+},b_{+}$ ($a_{-},b_{-}$), 
we simply group terms associated with $ D_{+} $ and $ b_{+} $ are grouped to $ G_{R,+}$ and perform the similar procedure for $G_{-}$.

We are now ready to take  the slow/fast limits as defined in the previous section. 
Noting
\begin{align}
&\,
\lim_{\textbf{s}}(\mp\Delta \e D_{\pm})=a_{\pm}\, ,
\qquad
\lim_{\textbf{f}}(q_{0} D_{\pm})= 0\, ,
\\
&\,\lim_{\textbf{s}}(\mp\Delta \e D_{\pm})=0\, ,
\qquad
\lim_{\textbf{f}}(q_{0} D_{\pm})= a_{\pm}\, ,
\end{align}
it immediately follows from Eq.~\eqref{G-i0j-final} and Eq.~\eqref{G-00j-final} that:
\begin{align}
\label{G-i0j-s}
\frac{i\partial}{\partial q_m}&\lim_{\textbf{s}}G^{i,0j}_{\pm}=
\frac{1}{2}\,\epsilon^{ijm}(a_{\pm}-b_{\pm}) +\frac{1}{2}\epsilon^{iml}v_{l}v^{j}(a_{\pm}+b_{\pm})
\\
\label{G-i0j-f}
&\frac{i\partial}{\partial q_m}\lim_{\textbf{f}}G^{i,0j}_{\pm}=
-\frac{1}{2}\,\epsilon^{ijm}b_{\pm}-\frac{1}{2}\,\epsilon^{ijl} v_{l}\,v^{m}a_{\pm}  +\frac{1}{2}\epsilon^{iml}v_{l}v^{j}b_{\pm}
\end{align}
Finally, we have:
\begin{align}
\label{G-i0j-s-f}
&\,
\k^{ijm}_{\pm}=\lim_{\textbf{s-f}}\frac{i\partial}{\partial q_m}G^{i,0j}_{\pm}=
\frac{1}{2}a_{\pm}(-\epsilon^{imj} +\epsilon^{iml}v_{l}v^{j}+\,\epsilon^{ijl} v_{l}\,v^{m})
\end{align}
Similarly, we have
\begin{align}
\label{G-00j-s-f}
&\lim_{\textbf{s-f}}\frac{i\partial}{\partial q_m}G^{0,0j}_{\pm}=
\mp\frac{1}{2}\epsilon^{jml}v_{l}\, a_{\pm}\, .
\end{align}
With Eqs.~\eqref{G-i0j-s-f}, \eqref{G-00j-s-f} at hand, we can extract the transport functions, as we shall do in the next section.

\section{Results and discussion}
\label{sec:dis}

\subsection{Collecting our results, and comparison to others}

In Sec.~\ref{sec:expansion}, 
we expand axial current phase space density $\sA^{\mu}(t,\vx,\vp)$ in terms of derivatives of hydrodynamics fields (see Eqs.~\eqref{f-exp},\eqref{A-exp}),
and derive a Kubo relation relating transport functions in this derivative expansion with the behavior of correlation function under appropriate limits, see Eqs.~\eqref{Kubo-c}, \eqref{kappa} for explicit expression.
In Sec.~\ref{sec:cal}, 
we performed a standard one-loop calculation and determined the behavior of the correlation function under those limits. 
Substituting Eqs.~\eqref{G-i0j-s-f}, \eqref{G-00j-s-f} int  Eqs.~\eqref{Kubo-c},\eqref{ct-o}, 
we obtain the desired expression for transport functions for fermions with a generic mass:
\begin{align}
\label{c-results}
&\, 
%(\tilde{c}_{\o,\pm})_{{\rm 1-loop}}=-\e_{u}\,\le(-n'_{\pm}(\e_{u})\ri)\,,
%\quad
(c_{\o,\pm})_{{\rm 1-loop}}=- (\tilde{c}_{\o,\pm})_{{\rm 1-loop}}=\e_{u}\,\le(-n'_{\pm}(\e_{u})\ri)\, ,
\quad
\\
\label{g-results}
&\,(g_{\o,\pm})_{{\rm 1-loop}}=0\, \qquad
 (g_{\s,\pm})_{{\rm 1-loop}}=(-v^{\a}v_{\a})\,\e_{u}\,\le(-n'_{\pm}(\e_{u})\ri)\, ,
\end{align}
where we introduce the notation $n'_{\pm}(\e_u)\equiv \pd n_{\pm}(\e_u)/\pd \e_{u}$. 
Since we are working in the rest-frame of the fluid in the previous sections, 
we have replaced $\e_{\vp}$ with $\e_{u}=p\cdot u$ and $\vv^{2}$ with $-(v^{\a}v_{\a})$  in Eqs.~\eqref{G-i0j-s-f}, \eqref{G-00j-s-f} to obtain results in the general fluid frame, i.e. Eq.~\eqref{c-results}.
In Ref.~\cite{Liu:2020dxg} by us, 
we computed $c_{T}, c_{\mu}$ through the same method, finding:
\begin{align}
\label{c-T} 
(c_{T,\pm})_{{\rm 1-loop}}= \e_{u}\le(-n'_{\pm}(\e_{u})\ri)\, ,
\qquad
(c_{\mu,\pm})_{{\rm 1-loop}}= \mp\b^{-1}\,\le(-n'_{\pm}(\e_{u})\ri)\, .
\end{align}
Placing expressions~\eqref{c-results},\eqref{g-results},\eqref{c-T} back to the derivative expansion Eqs.~\eqref{f-exp},\eqref{A-exp},
we have
\bes
\label{A-loop}
\begin{align}
(f_{A,\pm})_{{\rm 1-loop}}&=-\le(-n'_{\pm}(\e_{u})\ri)\, p_{\perp}\,\cdot \o\,
\\
(\sA^{\mu}_{\perp,\pm})_{{\rm 1-loop}}&= 
\e_{u}\,\le(-n'_{\pm}(\e_{u})\ri)
\Big\{
\,\o^{\mu}
+\epsilon^{\mu\nu\a\l} u_{\nu}\, \frac{p^{\perp}_{\a}}{\e_u}\,\le[\b^{-1}\,(\pd^{\perp}_{\l} \beta)\ri]\mp \,(\frac{1}{\b \e_u})\,\epsilon^{\mu\nu\a\l} u_{\nu} \frac{p^{\perp}_{\a}}{\e_u}\,\pd^{\perp}_{\l}(\b\mu)
\no \\
&\,+(-\frac{p^2_{\perp}}{\e^{2}_u})\,\epsilon^{\mu\nu\a\b}u_{\nu}Q\indices{_{\a}^{\rho}} \,\s_{\rho\beta }
\Big\}\, .
\end{align}
\ees

The  phase space density of axial current (or the average spin vector) in the presence of temperature gradient and vorticity has been studied extensively (see Ref.~\cite{Becattini:2020sww} for a review). 
To make comparison with previous studies, 
we use
\begin{align}
\label{Tvorticity-relation}
&\,\frac{1}{2}\epsilon^{\mu\nu\a\l}p_{\nu}\, \beta^{-1}\pd_{\a}(\b u_{\beta})
= \frac{1}{2}\epsilon^{\mu\nu\a\l}\le( \e_{u}u_{\nu}+p_{\nu,\perp}\ri)\,\le[u_{\l}\b^{-1}(\pd_{\a} \b) +\pd_{\a}u_{\l}\ri]
\no\\
&\, 
=\e_{u}\o^{\mu}+\frac{1}{2}\epsilon^{\mu\nu\a\l}p_{\perp,\nu}u_{\l}\b^{-1}(\pd_{\a}\b)+\frac{1}{2}\epsilon^{\mu\nu\a\l}p_{\perp,\nu}(u_{\a}Du_{\l}+\pd^{\perp}_{\a}u_{\l})\, ,
\no \\
&=\e_{u}\o^{\mu}+\epsilon^{\mu\nu\a\l}u_{\nu}p_{\perp,\a}\b^{-1}(\pd_{\l}\b)-u^{\mu}(p_{\perp}\cdot \o)\, ,
\end{align}
where from the second line to the third line, we have used the ideal hydrodynamic equation for a neutral fluid: $D u_{\a}=-\b^{-1}\pd_{\a} \b$ where $D\equiv u\cdot \pd$ and the relation:
\begin{align}
\label{relation}
\frac{1}{2}\epsilon^{\mu\nu\a\l}p_{\perp,\nu}\,\pd^{\perp}_{\a}u_{\l}=\frac{1}{2}\epsilon^{\mu\nu\a\b}p_{\perp,\nu}\,\epsilon_{\a\b\sigma\rho}u^{\sigma}\o^{\rho}
= -u^{\mu}(p_{\perp}\cdot \o)\, .  
\end{align}
Therefore we can recast Eq.~\eqref{A-loop} into the form:
\begin{align}
\label{A-compare}
(\sA^{\mu}_{\pm})_{{\rm loop}}=-n'_{\pm}(\e_{u})\,\le[
   \b^{-1} \tilde{\Omega}^{\mu\nu}_{T}p_{\nu}
    \mp \,(\frac{1}{\b \e_u})\,\epsilon^{\mu\nu\a\l} u_{\nu} p^{\perp}_{\a}\,\pd^{\perp}_{\l}(\b\mu)
    +(-\frac{p^2_{\perp}}{\e_u})\,\epsilon^{\mu\nu\a\b}u_{\nu}Q\indices{_{\a}^{\rho}} \,\s_{\rho\beta }
    \ri]
\end{align}
where (the dual of) thermal vorticity is given by $\tilde{\O}_{{\rm T}}^{\mu\nu}=(1/2)\epsilon^{\mu\nu\a\lambda}\pd_{\a}(\b u_{\l})$.
The first term in Eq.~\eqref{A-compare} matches with the description of thermal vorticity effects in Refs.~\cite{Becattini:2013fla,Fang:2016vpj}

Two remarks are due here:
\begin{itemize}
\item When shear strength and charge density gradient are non-zero, the thermal vorticity is not the only effect which contribute to spin polarization. 
The shear strength is generically comparable to vorticity in heavy-ion collisions. 
At beam energy scan energy at RHIC, the baryon density gradient is also sizable. 
\item Several alternative forms of axial Wigner function are proposed to describe the experiment data, see Refs.~\cite{Wu:2019eyi,Florkowski:2019voj} for examples. 
Those expression of course can be expressed in the form of Eqs.~\eqref{f-exp},\eqref{A-exp}. However, the corresponding transport functions are different from one-loop results Eqs.~\eqref{c-results}, \eqref{g-results}.

\end{itemize}

\subsection{Shear-induced polarization and magnetization current
	\label{sec:mag}
}

What is the physical origin of shear-induced quadrupole?
Why $g_{\o}$, which quantify the quadrupole induced by vorticity, is zero at one-loop order, although it does not have to be?
We shall address those questions by considering quantum kinetic theory. 

For simplicity, 
let us now consider a specific case that fermions under study is massless. 
Below, we shall only present discussion for fermions as that for anti-fermions is in parallel. 
According to chiral kinetic theory~\cite{Son:2012wh,Stephanov:2012ki} (see Ref~\cite{Mueller:2017arw,Mueller:2019gjj} for examples of recent developments), 
one can determine $\sA_{\pm}$ through the variation of the semi-classical action for Wely fermions~\cite{Chen:2014cla,Chen:2015gta},
yielding
\begin{align}
\label{A-CKT}
\sA_{+}(t,\vx;\vp)=\sum_{\l=\pm}\,\e_{\vp}\le[ s_{\l} \hat{\vp}\,f_{\l}(t,\vx;\vp) -\frac{\hat{\vp}}{2 |\vp|}\times \nabla f_{\l}(t,\vx;\vp)\ri]\, ,
\end{align}
where $\l=+,-$ accounts for right-handed and left-handed fermions respectively and where $(s_{L},s_{R})=(1,-1)$. 
Note for massless fermions, $\vv=\hat{\vp}$ and $\e_u=|\vp|$ in the fluid rest frame. 
The first term of Eq.~\eqref{A-CKT} arises from the fact that the spin of right-handed (left-handed) fermions is along (opposite to) the direction of its momentum. 
The second term is refered as to the magnetization current (MC) term $\nabla\times {\bm M}$, since the magnetization density ${\bm M}$ of chiral fermion $\l$ is related to magnetic moments of fermions: $ {\bm M}_{\l}=\int_{\vp} (\hat{p}/2|\vp|)\,f_{\l}\,$, 
and $\nabla\times {\bm M}$ is the magnetization current.

Let us consider the system in the presence of a slow varying velocity field ${\bm u}$, and examine the behavior of $f_{A}, \bm{\sA}_{+}$ according to Eq.~\eqref{A-CKT}. 
The local equilibrium distribution function becomes $n_{+}(\beta(\e_{\vp}-\vp\cdot {\bm u}-\Delta\e_{\o,\l}))$
where the energy shift due to spin-vorticity coupling is given by $\Delta \e_{\o,\l} =\frac{1}{2} s_{\l} {\bm \o}\cdot \hat{\vp}$\, . 
We first note that to the first order in gradient, $f_{A,+}$ is non-zero because of the energy shift:
\begin{align}
\label{f-A-shift}
f_{A,+}=\e_{\vp}\sum_{\l=\pm}n_{+}(\beta(\e_{\vp}-\vp\cdot {\bm u}-\Delta\e_{\o,\l}-\mu))=(-n'_{+})\e_{\vp}\,\bm{\o}\cdot\hat{\vp}\, ,
\end{align}
Next, 
we replace $f_{\l}$ with $n_{+}(\beta(\e_{\vp}-{\vp}\cdot {\bm u}-\Delta\e_{\o}))$ and expand Eq.~\eqref{A-CKT}  to the first order in gradient. 
 The first term in Eq.~\eqref{A-CKT}, induced by energy shift, becomes
\begin{align}
\label{e-shift-term}
&\,(\sA^{i}_{+})_{1}=(-n'_{+})\varepsilon_{\vp}
\le(\frac{1}{3}\o^{i}-Q^{ij}\o_{j}\ri)
\end{align}
while the second term, related to the MC term, can be written as
\begin{align}
\label{m-current-term}
(\sA^{i}_{+})_{2}&=
-(-n'_{+})\e_{\vp}\,\epsilon^{ilm}\hp_{l}\hp_{j}\pd_{m}u^{j}
\no \\
&=-(-n'_{+})\e_{\vp}\le(
-\frac{2}{3}\o^{i}-Q^{ik}\o_{k}+\epsilon^{ilm}Q_{lj}\s_{\,\,\,m}^{j}\ri)\, . 
\end{align}
Collecting both contributions,
we find
\begin{align}
\label{A-final}
\,\sA^{i}_{+}=&(-n'_{+})\e_{\vp}\Bigg[
\le(\frac{1}{3}\o^{i}-Q^{ij}\o_{j}\ri)-
\le(
-\frac{2}{3}\o^{i}-Q^{ij}\o_{j}+\epsilon^{ijk}Q_{jl}\s^{l}_{\,\,\,k}\ri)
\Bigg]
\no\\
=&(-n'_{+})\e_{\vp} \le(
\o^{i}-\epsilon^{ijk}Q_{jl}\s^{l}_{\,\,\,k}
\ri)\, .
\end{align}
Note,  Eq.~\eqref{A-final} agrees with Eq.~\eqref{A-exp} in the massless limit. 
To this point, we have assumed that $\b$ and $\mu$ are constant. 
It is a straightforward exercise to show that one could obtain the effects of temperature and chemical potential gradient from the magnetization current in Eq.~\eqref{A-CKT} following similar steps, and the results of doing so agree with Eq.~\eqref{A-exp} in massless limit as well. 
Since the one-loop calculations should be matched with the analysis based on kinetic theory,
we take this agreement as a nontrivial test of our calculation.

The above exercise makes clear the close relation between shear-induced polarization and the MC term. The latter is well-known in the context of quantum kinetic theory. 
One should be able to generalize the above discussion based on chiral kinetic theory for massless fermions to massive fermions. 
Indeed, Ref.~\cite{Hattori:2019ahi} shows that quantum kinetic equation and consequently $\sA^{\mu}_{\pm}$ behaves smoothly from massless limit to the massive case, 
see also Refs.~\cite{Liu:2020flb,Weickgenannt:2019dks,Weickgenannt:2020aaf} for related studies.

We close this section by discussing potential corrections to Eq.~\eqref{c-results} from higher loop effects along the lines of the above analysis.
Let us account for the high loop corrections to the energy shift due to spin-voriticity coupling by introducing a non-zero parameter $\g$, i.e., $\Delta \e_{\o,\l}=\frac{(1+\g)}{2}\vo\cdot\hp$. 
Further, we recover the g-factor $g$ dependence of the MC term which changes by $g/2$ (see for example Ref.~\cite{Kharzeev:2016sut}). 
Consequently, 
Eqs.~\eqref{f-A-shift} and \eqref{A-final} become, respectively, 
\begin{align}
\label{f-A-shift-2}
f_{A,+}&=(1+\gamma)(-n'_{+})\e_{\vp}\,\bm{\o}\cdot\hat{\vp}\, ,
\\
\label{A-final-2}
\,\sA^{i}_{+}&=(-n'_{+})\e_{\vp}\Bigg[
(1+\gamma)\le(\frac{1}{3}\o^{i}-Q^{ij}\o_{j}\ri)-
\frac{g}{2}\le(
-\frac{2}{3}\o^{i}-Q^{ij}\o_{j}+\epsilon^{ijk}Q_{jl}\s^{l}_{\,\,\,k}\ri)
\Bigg]
\no\\
=&(-n'_{+})\e_{\vp} \le[
\le(1+\frac{1}{3}(\g+g-2)\ri)\o^{i}-\le(1+\frac{g-2}{2}\ri)\epsilon^{ijk}Q_{jl}\s^{l}_{\,\,\,k}
+\frac{1}{2}\le((g-2)-\g\ri) Q^{ij}\o_{j}
\ri]\, .
\end{align}
Two comments is due here. 
First, we observe from \eqref{A-final-2} that $g_{\o}\propto \le((g-2)-\g\ri)$. 
Since $\g$ is generically independent of $g-2$, $g_{\o}$ might be non-zero beyond one-loop.
Second, we observe from Eqs.~\eqref{f-A-shift-2} and \eqref{A-final-2} that $c_{\o}\neq \tilde{c}_{\o}$, or more explicitly, $c_{\o}-\tilde{c}_{\o}\propto ((g-2)-2\g)$.
So, it would be interesting to evaluate radiative corrections to $g_{\o},c_{\o},\tilde{c}_{\o}$ as well other coefficients in future. 
We remark that in Dirac/Weyl semi-metals, the value of g-factor of the emergent chiral quasi-particles can significantly differ from $g=2$ due to a strong spin-orbit coupling (see Ref~\cite{Kharzeev:2016sut} and references therein).

\section{Summary and outlook
	\label{sec:outlook}
}

In the fireball created in heavy-ion collisions, 
the derivatives of hydrodynamic fields such as flow and temperature can be of the order $10$~MeV. 
Those derivatives might give rise to observable effects in $\Lambda$ and $\bar{\Lambda}$ spin polarization measurement under the current precision. 
In this work, we have expanded the axial Wigner function in phase space systematically to the first order in derivatives. 
To obtain the associated transport functions, we have derived a generalized Kubo relation, allowing for the computation of relevant transport functions using the standard thermal field theory techniques.  
The present one-loop calculation in massless limit agrees with those obtained using the chiral kinetic theory. 
However, 
our results can be systematically improved by including higher loop corrections.

\begin{figure}[t]
\centering
\includegraphics[width=0.4\textwidth]{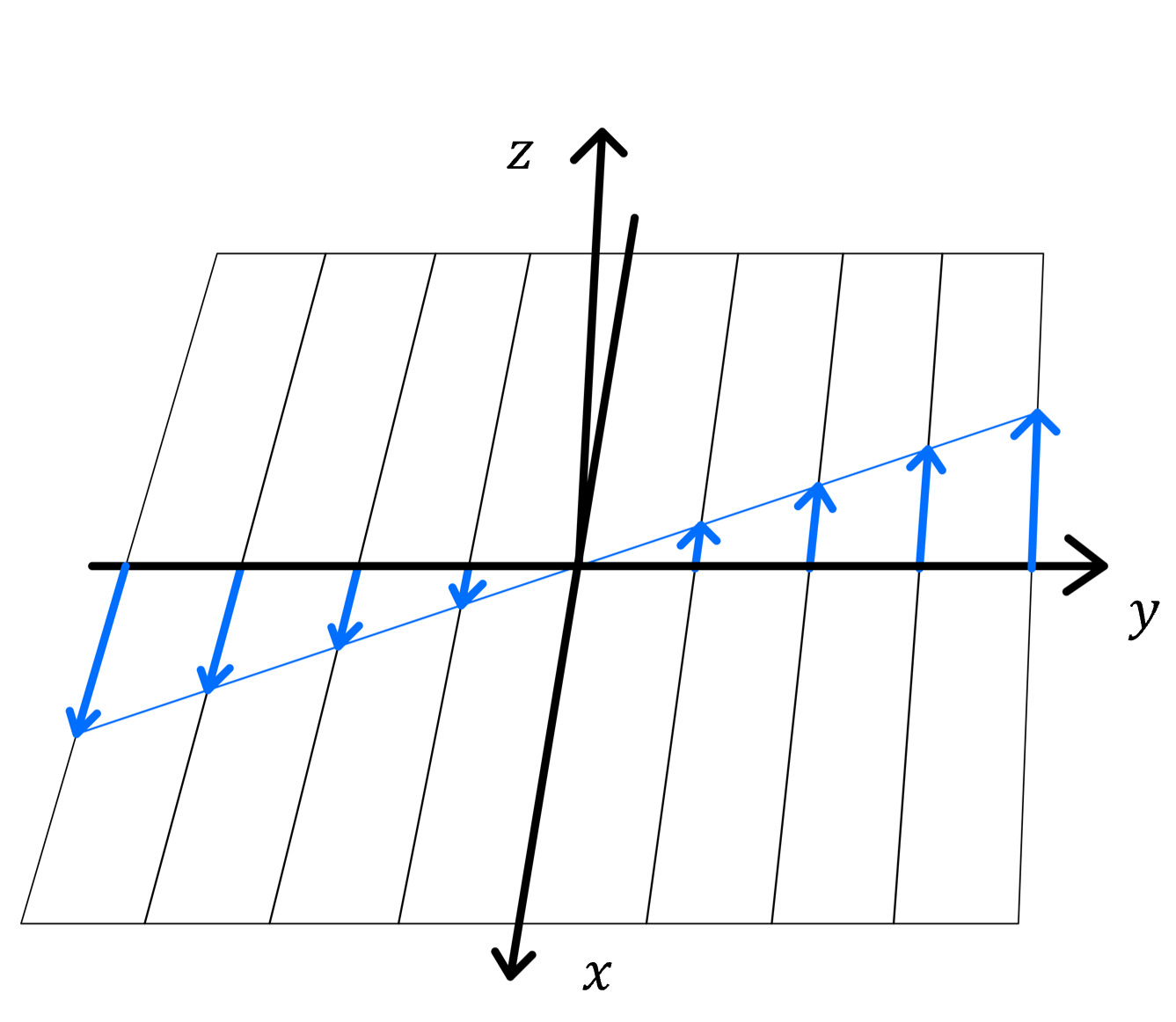}
\includegraphics[width=0.4\textwidth]{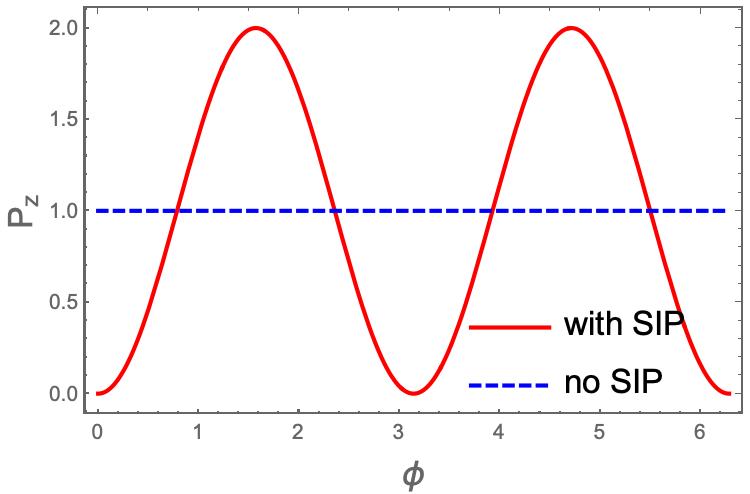}
\caption{
\label{fig:illustration}
An illustration of spin polarization density in phase space induced by a shear-flow profile. 
Left: we consider and plot a typical shear flow profile $\pd_{y}u_{x}=-\alpha$, where $\a$ is a constant. 
The corresponding vorticity and shear strength are $\s^{x}_y=\s^{y}_{x}=\a, \o^z=\a$. 
According to Eq.~\eqref{A-loop}, the spin polarization vector projected to $z$-direction is given by: $P^z\propto \sA^{z}\propto \a\le (1-\sin^{2}\theta\cos(2\phi)\ri)$. 
In the expression of $P^{z}$, the first term is due to vorticity-induced polarization (see the first term of Eq.~\eqref{A-loop}) and the second one is from "shear-induced polarization" (SIP), see the second line of Eq.~\eqref{A-loop} or Eq.~\eqref{v-tensor-pol} and the text. 
Right: the resulting (scaled) $P_z$ vs $\phi$ at $\theta=0$. 
We consider massless fermions for illustrative purpose and parametrize the direction of the single particle momentum in terms of polar and azimuthal angle $\hp=(\sin\theta\cos\phi, \theta\sin\phi,\cos\theta)$.
}
\end{figure}

The main message of our study is that derivatives of hydrodynamic fields can induce a rich pattern in axial Wigner function and hence spin polarization density in phase space.
In particular, we observe the shear strength (see Eq.~\eqref{shear}) can induce quadrupole pattern in the phase space, 
and this contribution is closely related to magnetic current. 
For illustration, 
let us take a flow profile $u^{\mu}=(1,\bm{u})$ with $|\bm{u}|\ll 1$.
Eq.~\eqref{g-results} then becomes
\begin{align}
\label{v-tensor-pol}
\sA^{i}_{\pm} =-(-\frac{\pd n_{\pm}}{\pd \e})\,(\frac{\vp}{\e_{\vp}})^{2} \epsilon^{ijk}Q_{jl}\s^{\,\,l}_{k} +\ldots\, ,
\end{align}
where explicitly
\begin{align}
Q^{ij}=\le(\hp^{i}\hp^{j}-\frac{1}{3}\d^{ij}\ri)\, ,
\qquad
\s_{i}^{\,\, j}=\frac{1}{2}\le(\pd_{i}u^{j}+\pd_{j}u^{i}+\frac{2}{3}({\bm \pd}\cdot {\bm u})\,\d_{i}^{\,\, j}\ri)\, . 
\end{align}
Let us further consider the familiar shear flow in which case both vorticity and shear strength are present. 
Fig.~\ref{fig:illustration}, which shows the resulting spin polarization in phase space, suggests a description of spin-polarization distribution would not be complete without including ``shear-induced polarization''.
We also note that the coupling between the quadrupole and vorticity (see the last term in Eq.~\eqref{A-exp}) 
might be non-zero beyond one-loop. 
This would potentially induce additional quadrupole pattern in the spin polarization distribution in the presence of vorticity.

In future, the derivative expansion supplemented with the one-loop calculation results, i.e., Eq.~\eqref{A-loop}, can be served as a reference basis for the interpretation of the experimental data. 
For this purpose, 
one needs to perform a quantitative study based on hydrodynamic derivatives obtained from the state-of-art hydrodynamics simulation, 
see also a recent study using the transport model~\cite{Liu:2019krs} which implements the contribution from the magnetization current. 
The results employing the realistic hydrodynamic modeling are reported in the accompanying paper~\cite{Fu:2021pok}.

\begin{acknowledgments}
	We are grateful to
	Baochi Fu,
	Longgang Pang,
	Koichi Hattori,
	Chun Shen,
	Shi Pu,
	Shu Lin,
	Shuzhe Shi,
	Naoki Yamamoto,
	Dilun Yang,
	Lixing Yang
	for helpful conversations.
	In particular, we thank Yoshimasa Hidaka, Feng Li, Yifeng Sun, Shu Lin, Ho-Ung Yee and  Dilun Yang for comments on the draft. 
	This work was supported by the Strategic Priority Research Program of Chinese Academy of Sciences, Grant No. XDB34000000.
\end{acknowledgments}

\appendix

\section{The details in the gradient expansion
\label{sec:gradient-more}
}

We here provide further details on obtaining the gradient expansion in Eqs.~\eqref{f-exp} and \eqref{A-exp}. 
Our strategy is to first write down all possible vectors which can be formed by each of the derivatives listed in Eq.~\eqref{gradient} together with $u^{\mu}, p^{\mu}_{\perp}, Q^{\mu\nu}, \epsilon^{\mu\nu\a\b}$ , and then examine how the results transform under $\CT$ and $\CP$. 
For simplicity, 
we shall consider a neutral fluid and hence do not include the contribution from $\pd^{\mu}_{\perp}(\b\mu)$ for the moment. 
There are fourteen possible vectors:
\begin{align}
\label{W-list}
W^{\mu}_{1}&=\theta u^{\mu}\, , \qquad 
W^{\mu}_{2}=\theta p^{\mu}_{\perp}\, , 
\\
W^{\mu}_{3}&=(\o\cdot p) u^{\mu}\, , \qquad
W^{\mu}_{4}=\omega^{\mu}\, , \qquad
W^{\mu}_{5}=Q^{\mu\nu}\omega_{\nu}\, , \qquad
W^{\mu}_{6}=\e^{\mu\nu\a\b}u_{\nu}p_{\a}\o_{\b}
\\
W^{\mu}_{7}&=(a\cdot p) u^{\mu}\, , \qquad
W^{\mu}_{8}=a^{\mu}\, , \qquad
W^{\mu}_{9}=Q^{\mu\nu}\, a_{\mu}\, , \qquad
W^{\mu}_{10}=\e^{\mu\nu\a\b}u_{\nu}p_{\a}a_{\b}\, , \qquad
\\
W^{\mu}_{11}&=(a_{\s}\cdot p) u^{\mu}\, , \qquad
W^{\mu}_{12}=a^{\mu}_{\sigma}\, , \qquad
W^{\mu}_{13}=Q^{\mu\nu}\, a_{\mu,\sigma}\, , \qquad
W^{\mu}_{14}=\e^{\mu\nu\a\b}u_{\nu}p_{\a}a_{\b,\sigma}\, . 
\end{align} 
where we have defined
\begin{align}
    a^{\mu}\equiv \pd^{\perp}_{\mu}\log\beta\, , 
    \qquad
    a^{\mu}_{\s}\equiv \sigma^{\mu\nu}p_{\nu} .
\end{align}
Using the Table.~\ref{Table-1} where we list transformations of various tensors under $\CP$ and $\CT$, it is easy to check:
\begin{align}
\label{W-not}
    \CT: \,\, W^{\mu}_{b}\to (-W^{0}_{b}, W^{i}_{b})\qquad
    \CP: \,\, W^{\mu}_{b}\to (W^{0}_{b}, -W^{i}_{b})\qquad
    b=1,2,6,7,8,9,11,12,13\, .
\end{align}
while $\sA^{\mu}$ transforms as:
\begin{align}
    \CT: \,\, \sA^{\mu}\to (\sA^{0}, -\sA^{i}_{b})\qquad
    \CP: \,\, \sA^{\mu}_{b}\to (-\sA^{0}, \sA^{i}_{b})\qquad
\end{align}
Therefore $W_{b}$ listed in Eq.~\eqref{W-not} can not enter in the derivative expansion of $\sA$ for the parity-even background under consideration since they transform differently from $\sA$ under $\CP$. 
The remaining five vectors, $W_{3},W_{4},W_{5},W_{10},W_{14}$,  transform in the same way as $\sA^{\mu}$ under $\CP$ and hence are allowed by symmetry. 
Furthermore, one can verify using Table.~\ref{Table-1} that $W^{\mu}_{3},W^{\mu}_{4},W^{\mu}_{5},W^{\mu}_{10},W^{\mu}_{14}$ behave in the same way as $\sA^{\mu}$ under $\CT$.
So, 
the corresponding expansion coefficients, defined in Eqs.~\eqref{f-exp}, \eqref{A-exp}, are $\CT$-even, meaning effects described by those equations, including shear-induce polarization, are \textit{non-dissipative}. 
Extending the above analysis to the charged fluid, we arrive at Eqs.~\eqref{f-exp}, \eqref{A-exp}.  
Note if we were studying a fluid with background axial charge density, terms listed in Eq.~\eqref{W-not} would be allowed. 
For example, the effect associated with $W_{6}$ is recently discussed in Ref.~\cite{Hou:2020mqp}.

\begin{table}[ht]
\begin{center}
\begin{tabular}{ |p{2cm}||p{3cm}|p{2cm}|  }
%  \hline
%  \multicolumn{3}{|c|}{Discrete transformations} \\
  \hline
 & $\CT$ &  $\CP$ \\
  \hline
 $\pd_\mu$   & $(-\pd_0,  \pd_i)$    & $(\pd_0,  -\pd_i)$   \\
 $u^\mu$   & $(u^0, - u^i)$    & $(u^0,  -u^i)$    \\
 $\o^\mu$   & $(\o^0, - \o^i)$    & $(-\o^0,  \o^i)$    \\
 $p^{\mu}_{\perp}$ &   $(p^{0}_{\perp},-p^{i}_{\perp})$     &  $(p^{0}_{\perp},-p^{i}_{\perp})$   \\
  $a^{\mu}$ &   $(-a^{0},a^{i})$   &  $(a^{0},-a^{i})$   \\
 $a^{\mu}_{\sigma}$ & $(-a^{0}_{\s},a^{i}_{\s})$ & $(a^{0}_{\s},-a^{i}_{\s})$ \\
  $\theta$ &    $-\theta$   &  $\theta$ \\
 \hline
\end{tabular}
\caption{
\label{Table-1}
Transformations of various tensors under $\CT$ and $\CP$.}
\end{center}
\end{table}

\bibliographystyle{JHEP}
\bibliography{refs}

\providecommand{\href}[2]{#2}\begingroup\raggedright\begin{thebibliography}{10}

\bibitem{Liang:2004ph}
Z.-T. Liang and X.-N. Wang, \emph{{Globally polarized quark-gluon plasma in
  non-central A+A collisions}},
  \href{https://doi.org/10.1103/PhysRevLett.94.102301,
  10.1103/PhysRevLett.96.039901}{\emph{Phys. Rev. Lett.} {\bfseries 94} (2005)
  102301}.

\bibitem{Becattini:2013fla}
F.~Becattini, V.~Chandra, L.~Del~Zanna and E.~Grossi, \emph{{Relativistic
  distribution function for particles with spin at local thermodynamical
  equilibrium}}, \href{https://doi.org/10.1016/j.aop.2013.07.004}{\emph{Annals
  Phys.} {\bfseries 338} (2013) 32}.

\bibitem{Becattini:2020sww}
F.~Becattini, \emph{{Polarization in relativistic fluids: a quantum field
  theoretical derivation}},  4, 2020,
  \href{https://arxiv.org/abs/2004.04050}{{\ttfamily 2004.04050}}.

\bibitem{Florkowski:2017ruc}
W.~Florkowski, B.~Friman, A.~Jaiswal and E.~Speranza, \emph{{Relativistic fluid
  dynamics with spin}},
  \href{https://doi.org/10.1103/PhysRevC.97.041901}{\emph{Phys. Rev. C}
  {\bfseries 97} (2018) 041901}
  [\href{https://arxiv.org/abs/1705.00587}{{\ttfamily 1705.00587}}].

\bibitem{Hattori:2019lfp}
K.~Hattori, M.~Hongo, X.-G. Huang, M.~Matsuo and H.~Taya, \emph{{Fate of spin
  polarization in a relativistic fluid: An entropy-current analysis}},
  \href{https://doi.org/10.1016/j.physletb.2019.05.040}{\emph{Phys. Lett. B}
  {\bfseries 795} (2019) 100}
  [\href{https://arxiv.org/abs/1901.06615}{{\ttfamily 1901.06615}}].

\bibitem{Liu:2019krs}
S.~Y.~F. Liu, Y.~Sun and C.~M. Ko, \emph{{Spin Polarizations in a Covariant
  Angular-Momentum-Conserved Chiral Transport Model}},
  \href{https://doi.org/10.1103/PhysRevLett.125.062301}{\emph{Phys. Rev. Lett.}
  {\bfseries 125} (2020) 062301}
  [\href{https://arxiv.org/abs/1910.06774}{{\ttfamily 1910.06774}}].

\bibitem{Fukushima:2020ucl}
K.~Fukushima and S.~Pu, \emph{{Spin Hydrodynamics and Symmetric Energy-Momentum
  Tensors -- A current induced by the spin vorticity --}},
  \href{https://arxiv.org/abs/2010.01608}{{\ttfamily 2010.01608}}.

\bibitem{Shi:2020htn}
S.~Shi, C.~Gale and S.~Jeon, \emph{{Relativistic Viscous Spin Hydrodynamics
  from Chiral Kinetic Theory}},
  \href{https://arxiv.org/abs/2008.08618}{{\ttfamily 2008.08618}}.

\bibitem{Li:2020eon}
S.~Li, M.~A. Stephanov and H.-U. Yee, \emph{{Non-dissipative second-order
  transport, spin, and pseudo-gauge transformations in hydrodynamics}},
  \href{https://arxiv.org/abs/2011.12318}{{\ttfamily 2011.12318}}.

\bibitem{Singh:2020rht}
R.~Singh, G.~Sophys and R.~Ryblewski, \emph{{Spin polarization dynamics in the
  Gubser-expanding background}},
  \href{https://arxiv.org/abs/2011.14907}{{\ttfamily 2011.14907}}.

\bibitem{STAR:2017ckg}
{\scshape STAR} collaboration, L.~Adamczyk et~al., \emph{{Global $\Lambda$
  hyperon polarization in nuclear collisions: evidence for the most vortical
  fluid}}, \href{https://doi.org/10.1038/nature23004}{\emph{Nature} {\bfseries
  548} (2017) 62}.

\bibitem{Adam:2018ivw}
{\scshape STAR} collaboration, J.~Adam et~al., \emph{{Global polarization of
  $\Lambda$ hyperons in Au+Au collisions at $\sqrt{s_{_{NN}}}$ = 200 GeV}},
  \href{https://doi.org/10.1103/PhysRevC.98.014910}{\emph{Phys. Rev. C}
  {\bfseries 98} (2018) 014910}
  [\href{https://arxiv.org/abs/1805.04400}{{\ttfamily 1805.04400}}].

\bibitem{Niida:2018hfw}
{\scshape STAR} collaboration, T.~Niida, \emph{{Global and local polarization
  of $\Lambda$ hyperons in Au+Au collisions at 200 GeV from STAR}},
  {\emph{Nucl. Phys. A} {\bfseries 982} (2019) 511}.

\bibitem{Adam:2019srw}
{\scshape STAR} collaboration, J.~Adam et~al., \emph{{Polarization of $\Lambda$
  ($\bar{\Lambda}$) hyperons along the beam direction in Au+Au collisions at
  $\sqrt{s_{_{NN}}}$ = 200 GeV}},
  \href{https://doi.org/10.1103/PhysRevLett.123.132301}{\emph{Phys. Rev. Lett.}
  {\bfseries 123} (2019) 132301}
  [\href{https://arxiv.org/abs/1905.11917}{{\ttfamily 1905.11917}}].

\bibitem{Becattini:2017gcx}
F.~Becattini and I.~Karpenko, \emph{{Collective Longitudinal Polarization in
  Relativistic Heavy-Ion Collisions at Very High Energy}},
  \href{https://doi.org/10.1103/PhysRevLett.120.012302}{\emph{Phys. Rev. Lett.}
  {\bfseries 120} (2018) 012302}.

\bibitem{Xia:2018tes}
X.-L. Xia, H.~Li, Z.-B. Tang and Q.~Wang, \emph{{Probing vorticity structure in
  heavy-ion collisions by local $\Lambda$ polarization}},
  \href{https://doi.org/10.1103/PhysRevC.98.024905}{\emph{Phys. Rev. C}
  {\bfseries 98} (2018) 024905}
  [\href{https://arxiv.org/abs/1803.00867}{{\ttfamily 1803.00867}}].

\bibitem{Chen:2014cla}
J.-Y. Chen, D.~T. Son, M.~A. Stephanov, H.-U. Yee and Y.~Yin, \emph{{Lorentz
  Invariance in Chiral Kinetic Theory}}, {\emph{Phys. Rev. Lett.} {\bfseries
  113} (2014) 182302}.

\bibitem{Chen:2015gta}
J.-Y. Chen, D.~T. Son and M.~A. Stephanov, \emph{{Collisions in Chiral Kinetic
  Theory}}, \href{https://doi.org/10.1103/PhysRevLett.115.021601}{\emph{Phys.
  Rev. Lett.} {\bfseries 115} (2015) 021601}.

\bibitem{Hattori:2019ahi}
K.~Hattori, Y.~Hidaka and D.-L. Yang, \emph{{Axial Kinetic Theory and Spin
  Transport for Fermions with Arbitrary Mass}},
  \href{https://doi.org/10.1103/PhysRevD.100.096011}{\emph{Phys. Rev. D}
  {\bfseries 100} (2019) 096011}
  [\href{https://arxiv.org/abs/1903.01653}{{\ttfamily 1903.01653}}].

\bibitem{Crooker_2005}
S.~A. Crooker and D.~L. Smith, \emph{Imaging spin flows in semiconductors
  subject to electric, magnetic, and strain fields},
  \href{https://doi.org/10.1103/physrevlett.94.236601}{\emph{Physical Review
  Letters} {\bfseries 94} (2005) }.

\bibitem{2005PhRvL..95j7203M}
A.~G. {Mal'Shukov}, C.~S. {Tang}, C.~S. {Chu} and K.~A. {Chao},
  \emph{{Strain-Induced Coupling of Spin Current to Nanomechanical
  Oscillations}},
  \href{https://doi.org/10.1103/PhysRevLett.95.107203}{\emph{Phys.Rev.Lett}
  {\bfseries 95} (2005) 107203}
  [\href{https://arxiv.org/abs/cond-mat/0504773}{{\ttfamily
  cond-mat/0504773}}].

\bibitem{Kharzeev:2016sut}
D.~E. Kharzeev, M.~A. Stephanov and H.-U. Yee, \emph{{Anatomy of chiral
  magnetic effect in and out of equilibrium}},
  \href{https://doi.org/10.1103/PhysRevD.95.051901}{\emph{Phys. Rev. D}
  {\bfseries 95} (2017) 051901}
  [\href{https://arxiv.org/abs/1612.01674}{{\ttfamily 1612.01674}}].

\bibitem{Liu:2020dxg}
S.~Y.~F. Liu and Y.~Yin, \emph{{Spin Hall effect in heavy ion collisions}},
  \href{https://arxiv.org/abs/2006.12421}{{\ttfamily 2006.12421}}.

\bibitem{Luttinger1964}
J.~M. Luttinger, \emph{Theory of thermal transport coefficients},
  \href{https://doi.org/10.1103/PhysRev.135.A1505}{\emph{Phys. Rev.} {\bfseries
  135} (1964) A1505}.

\bibitem{Liu:2020flb}
Y.-C. Liu, K.~Mameda and X.-G. Huang, \emph{{Covariant Spin Kinetic Theory I:
  Collisionless Limit}},
  \href{https://doi.org/10.1088/1674-1137/44/9/094101}{\emph{Chin. Phys. C}
  {\bfseries 44} (2020) 094101}
  [\href{https://arxiv.org/abs/2002.03753}{{\ttfamily 2002.03753}}].

\bibitem{Hayata:2020sqz}
T.~Hayata, Y.~Hidaka and K.~Mameda, \emph{{Second order chiral kinetic theory
  under gravity and antiparallel charge-energy flow}},
  \href{https://arxiv.org/abs/2012.12494}{{\ttfamily 2012.12494}}.

\bibitem{Landsteiner:2011iq}
K.~Landsteiner, E.~Megias, L.~Melgar and F.~Pena-Benitez, \emph{{Holographic
  Gravitational Anomaly and Chiral Vortical Effect}},
  \href{https://doi.org/10.1007/JHEP09(2011)121}{\emph{JHEP} {\bfseries 09}
  (2011) 121} [\href{https://arxiv.org/abs/1107.0368}{{\ttfamily 1107.0368}}].

\bibitem{Lin:2018aon}
S.~Lin and L.~Yang, \emph{{Mass correction to chiral vortical effect and chiral
  separation effect}},
  \href{https://doi.org/10.1103/PhysRevD.98.114022}{\emph{Phys. Rev.}
  {\bfseries D98} (2018) 114022}
  [\href{https://arxiv.org/abs/1810.02979}{{\ttfamily 1810.02979}}].

\bibitem{Fang:2016vpj}
R.-h. Fang, L.-g. Pang, Q.~Wang and X.-n. Wang, \emph{{Polarization of massive
  fermions in a vortical fluid}},
  \href{https://doi.org/10.1103/PhysRevC.94.024904}{\emph{Phys. Rev.}
  {\bfseries C94} (2016) 024904}
  [\href{https://arxiv.org/abs/1604.04036}{{\ttfamily 1604.04036}}].

\bibitem{Wu:2019eyi}
H.-Z. Wu, L.-G. Pang, X.-G. Huang and Q.~Wang, \emph{{Local spin polarization
  in high energy heavy ion collisions}},
  \href{https://doi.org/10.1103/PhysRevResearch.1.033058}{\emph{Phys. Rev.
  Research.} {\bfseries 1} (2019) 033058}
  [\href{https://arxiv.org/abs/1906.09385}{{\ttfamily 1906.09385}}].

\bibitem{Florkowski:2019voj}
W.~Florkowski, A.~Kumar, R.~Ryblewski and A.~Mazeliauskas, \emph{{Longitudinal
  spin polarization in a thermal model}},
  \href{https://doi.org/10.1103/PhysRevC.100.054907}{\emph{Phys. Rev. C}
  {\bfseries 100} (2019) 054907}
  [\href{https://arxiv.org/abs/1904.00002}{{\ttfamily 1904.00002}}].

\bibitem{Son:2012wh}
D.~T. Son and N.~Yamamoto, \emph{{Berry Curvature, Triangle Anomalies, and the
  Chiral Magnetic Effect in Fermi Liquids}},
  \href{https://doi.org/10.1103/PhysRevLett.109.181602}{\emph{Phys. Rev. Lett.}
  {\bfseries 109} (2012) 181602}
  [\href{https://arxiv.org/abs/1203.2697}{{\ttfamily 1203.2697}}].

\bibitem{Stephanov:2012ki}
M.~Stephanov and Y.~Yin, \emph{{Chiral Kinetic Theory}},
  \href{https://doi.org/10.1103/PhysRevLett.109.162001}{\emph{Phys. Rev. Lett.}
  {\bfseries 109} (2012) 162001}
  [\href{https://arxiv.org/abs/1207.0747}{{\ttfamily 1207.0747}}].

\bibitem{Mueller:2017arw}
N.~Mueller and R.~Venugopalan, \emph{{Worldline construction of a covariant
  chiral kinetic theory}},
  \href{https://doi.org/10.1103/PhysRevD.96.016023}{\emph{Phys. Rev. D}
  {\bfseries 96} (2017) 016023}
  [\href{https://arxiv.org/abs/1702.01233}{{\ttfamily 1702.01233}}].

\bibitem{Mueller:2019gjj}
N.~Mueller and R.~Venugopalan, \emph{{Constructing phase space distributions
  with internal symmetries}},
  \href{https://doi.org/10.1103/PhysRevD.99.056003}{\emph{Phys. Rev. D}
  {\bfseries 99} (2019) 056003}
  [\href{https://arxiv.org/abs/1901.10492}{{\ttfamily 1901.10492}}].

\bibitem{Weickgenannt:2019dks}
N.~Weickgenannt, X.-L. Sheng, E.~Speranza, Q.~Wang and D.~H. Rischke,
  \emph{{Kinetic theory for massive spin-1/2 particles from the Wigner-function
  formalism}}, \href{https://doi.org/10.1103/PhysRevD.100.056018}{\emph{Phys.
  Rev. D} {\bfseries 100} (2019) 056018}
  [\href{https://arxiv.org/abs/1902.06513}{{\ttfamily 1902.06513}}].

\bibitem{Weickgenannt:2020aaf}
N.~Weickgenannt, E.~Speranza, X.-l. Sheng, Q.~Wang and D.~H. Rischke,
  \emph{{Generating spin polarization from vorticity through nonlocal
  collisions}},  \href{https://arxiv.org/abs/2005.01506}{{\ttfamily
  2005.01506}}.

\bibitem{Fu:2021pok}
B.~Fu, S.~Y.~F. Liu, L.~Pang, H.~Song and Y.~Yin, \emph{{Shear-induced spin
  polarization in heavy-ion collisions}},
  \href{https://arxiv.org/abs/2103.10403}{{\ttfamily 2103.10403}}.

\bibitem{Hou:2020mqp}
D.~Hou and S.~Lin, \emph{{Polarization Rotation of Chiral Fermions in Vortical
  Fluid}},  \href{https://arxiv.org/abs/2008.03862}{{\ttfamily 2008.03862}}.

\end{thebibliography}\endgroup

\end{document}